\documentclass[10pt,preprint]{aastex}
\usepackage{graphics,epsf}
\usepackage{amsmath}                
\usepackage{amsfonts}               
\usepackage{amssymb}                
\usepackage{epsfig}                 
\usepackage[tight]{subfigure}
\usepackage{multirow}
\usepackage{longtable}

\def \cm{~\rm{cm}}
\def \s{~\rm{s}}
\def \km{~\rm{km}}

\def \erg{~\rm{erg}}

\def \yr{~\rm{yr}}
\def \Myr{~\rm{Myr}}

\def \kpc{~\rm{kpc}}

\def \keV{~\rm{keV}}

\begin{document}
\title{INFLATING A CHAIN OF X-RAY DEFICIENT BUBBLES BY A SINGLE JET ACTIVITY EPISODE}

\author{Michael Refaelovich\altaffilmark{1} and Noam Soker\altaffilmark{1}}

\altaffiltext{1}{Department of Physics, Technion -- Israel Institute of
Technology, Haifa 32000 Israel; mishar@tx.technion.ac.il; soker@physics.technion.ac.il.}

\begin{abstract}
We show that a continuous jet with time-independent launching properties can inflate a chain of close and overlapping X-ray deficient bubbles.
Using the numerical code PLUTO we run 2.5D {{{(i.e. spherical coordinate system with cylindrical symmetry)}}} hydrodynamic simulations and study the interaction of the jets with the intra-cluster medium (ICM).  A key process is vortex fragmentation due to several mechanisms, including vortex-shedding and Kelvin-Helmholtz (KH) instabilities.
Our results can account for the structure of two opposite chains of close bubbles
as observed in the galaxy cluster Hydra A.
Our results imply that the presence of multiple pairs of bubbles does not necessarily imply several
jet-launching episodes.
This finding might have implications to feedback mechanisms operating by jets.
\end{abstract}

\section{INTRODUCTION}
\label{sec:intro}

Bubbles (cavities) devoid of X-ray emission, mostly as opposite pairs, are
observed in a large fraction of cooling flow clusters and groups of
galaxies, as well as in cooling flow elliptical galaxies (e.g., \citealt{Dong2010}).
Examples include Hydra A \citep{Wise2007}, RBS797
\citep{Doria2012, Cavagnolo2011}, Perseus \citep{Fabian2011}, Abell 2052 \citep{Blanton2011},
NGC 6338 \citep{Pandge2012}, NGC 5044 \citep{David2011}, NGC 4636
\citep{Baldi2009}, NGC 5044 \citep{David2009}, HCG 62 \citep{Gitti2010} and NGC
5044 \citep{Gastaldello2009}.
These bubbles are inflated by jets launched from the central active galactic nuclei (AGN),
as evident by the radio emission that fills most bubbles. These jets and
cavities heat the intracluster medium (ICM; e.g., \citealt{OSullivan2011,
Gaspari2012a, Gaspari2012b, Birzan2011, Gitti2012} for recent papers and
references therein), and maintain a negative feedback mechanism with the
cooling gas \citep{Binney1995,Farage2012}.

Wide bubbles very close to the origin of the jets (the AGN), e.g., as in Abell
2052, that are termed `fat bubbles', can be inflated by jets that do not
penetrate through the ICM. Instead, they deposit their energy relatively close
to their origin and inflate the fat bubbles. Slow massive wide (SMW) jets can
inflate the fat bubbles that are observed in many cooling flows, in clusters, groups of
galaxies, and in elliptical galaxies \citep{Sternberg2007}. The same basic
physics that prevents wide jets from penetrating through the ICM holds for
precessing jets \citep{Sternberg2008a, Falceta-Goncalves2010} or a relative
motion of the jets to the medium \citep{Bruggen2007, Soker2009, Morsony2010, Mendygral2012}.
If the jet penetrate to a too large distance, then no bubbles are formed,
while in intermediate cases elongated and/or detached from the center
bubbles are formed (e.g., \citealt{Basson2003, Omma2004, Heinz2006, Vernaleo2006, AlouaniBibi2007, Sternberg2007, ONeill2010, Mendygral2011}).

Vortices inside the bubbles and in their surroundings play major roles in the
formation of bubbles, their evolution, and their interaction with the ICM.
\cite{Omma2004} find that a turbulent vortex trails each cavity, and that this
vortex contains a significant quantity of entrained and uplifted material
(also \citealt{Roediger2007}). Jet-excited shocks that interact with older bubbles can
excite vortices that dissipate energy to the ICM \citep{Friedman2012}. The
vortices cause semi-periodic changes in the bubble properties, such as its
boundary. This causes a single bubble to excite several sound waves \citep{Sternberg2009}.
Vortices play a major role in mixing the shocked jets' material with the ICM, hence heating the ICM gas
\citep{Gilkis2012}. The role of vortices in the interaction of the jet with the ICM was studied before (
e.g., \citealt{Norman1996, Mizuta2004}). However, no special emphasize on bubbles formation was done.

In some cases two opposite chains of bubbles that are close to each other, and even overlap, are observed
as in Hydra A \citep{Wise2007}, and in two bubbles in the galaxy group NGC 5813 \citep{Randall2011}. These chains were usually attributed to several episodes of jet activity. Using a 2.5D hydrodynamical code (section \ref{sec:numerical}) we study
the formation and evolution of chain of bubbles from one continuous jet activity episode (section \ref{sec:results}). Our short summary is in section \ref{sec:summary}.

\section{NUMERICAL SET UP}
\label{sec:numerical}

We use the multidimensional hydrodynamic PLUTO code \citep{Mignone2007} for all our simulations.
The simulations are 2.5D in the sense that we use a spherical coordinate system, but impose
cylindrical symmetry, hence calculating the flow with a 2D polar grid $(r,\theta)$.
There is also a mirror symmetry on the plane $\theta=\pi/2$.
There are 256 divisions in the azimuthal direction ($0 \leq \theta\leq \frac{\pi}{2}$) and 257 divisions in the
radial direction, geometrically stretched in the range ($1\leq r \leq 400 \kpc$).
Our flow is non-relativistic.
Gravity is included, but as the simulations last for a time shorter than the radiative cooling time we neglect radiative cooling.
The exact scheme that we use includes {{{ third-order Runge-Kutta}}} time-stepping {{{ integration (two predictors)}}}, parabolic interpolation and no dimensional splitting.

The boundary conditions are reflective on both 'angular' boundaries: the symmetry axis $\theta=0$
and the equatorial plane $\theta=\pi/2$, and an outflow condition on the far radius (outer radial boundary).
The jet is injected by using the boundary conditions at $r=1 \kpc$.
The jet is injected within an angle $0\leq\theta\leq\alpha$ -- where $\alpha$ is the half-opening angle.
At other angles on the inner radius we impose reflective boundary conditions.

The initial density profile is based on that of Hydra A as reported by \cite{McNamara2000}.
In the inner $10 \kpc$ we approximate the profile by a constant density of
$n_{e}=0.06 \cm^{-3}$, where $n_{e}$ is the electron density, while at $r>10 \kpc$ the
density drops as $n_e(r) \propto r^{-1}$.
The initial temperature of the gas was constant across the entire domain  $T_0=3.5 \keV$.
The gravity is calculated from the initial pressure and density profiles by assuming
hydrostatic equilibrium $g=-\frac{1}{\rho_0(r)}\frac{ d P_0(r)}{d r}$.

We performed many simulations spanning a large volume of the parameters space, e.g.,
various opening angles, jet powers, and jet velocities.
The standard case is taken to have the following parameters:
Total two jets power of $P_{2j} = 4 \times 10^{45} \erg \s^{-1}$, a jet Mach number of $15$ relative to the ambient medium,
a jet velocity of $v_j = 1.33 \times 10^{4} \km \s^{-1}$, hence the mass outflow from the two opposite jets (here we calculate only one jet) is $\dot M_{2j} =  70 M_\odot \yr^{-1}$, and a half-opening angle of $\alpha=30^{\circ}$.

\section{RESULTS}
\label{sec:results}
\subsection{Morphology}
\label{subsec:morphology}
\begin{figure}
  \centering
  \subfigure[][]{\label{subfigure:stdt}\includegraphics*[width=0.4\textwidth]{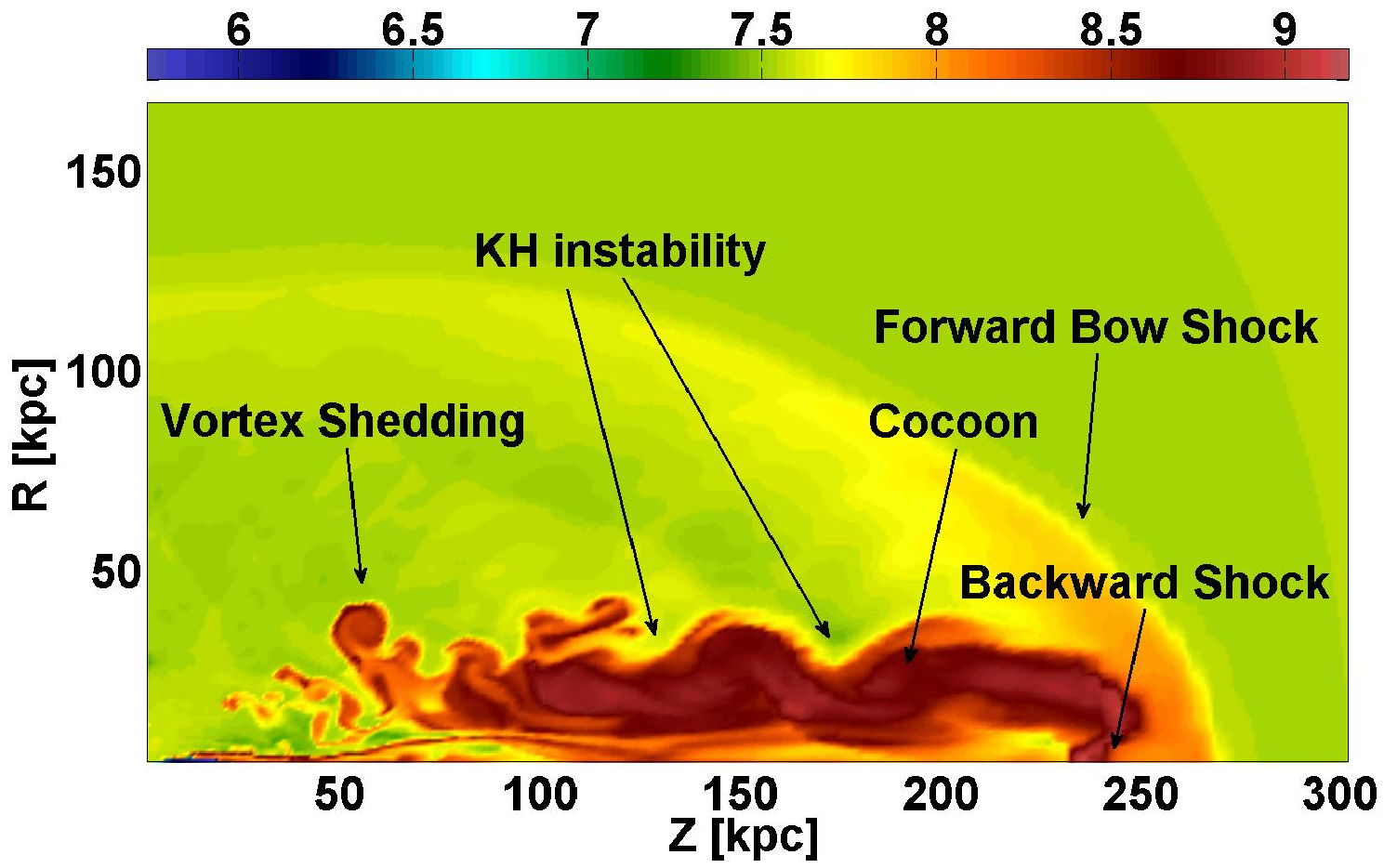}}
    \subfigure[][]{\label{subfigure:stdne}\includegraphics*[width=0.4\textwidth]{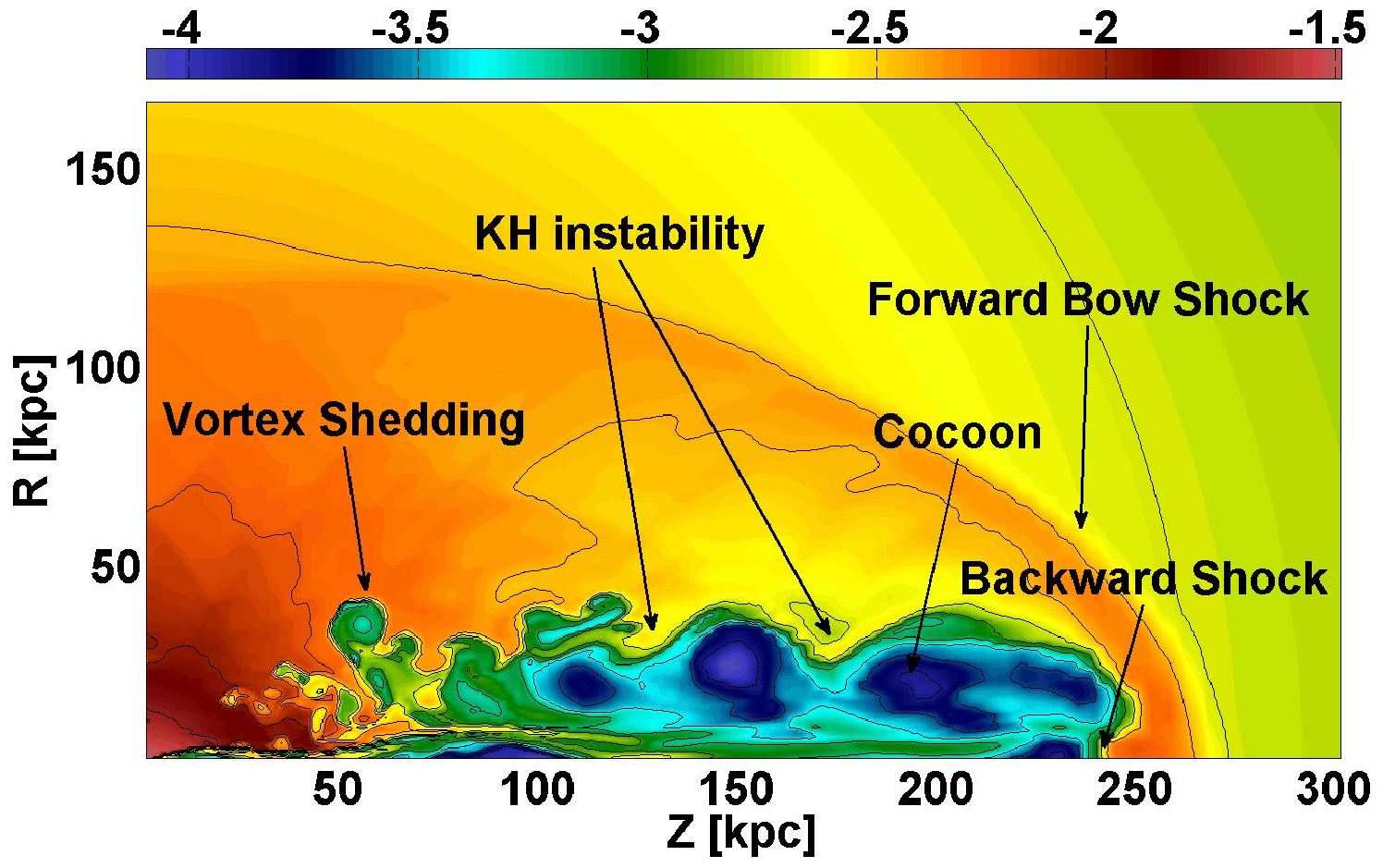}}\\
   \subfigure[][] {\label{subfigure:stdvortices}\includegraphics*[width=0.8\textwidth]{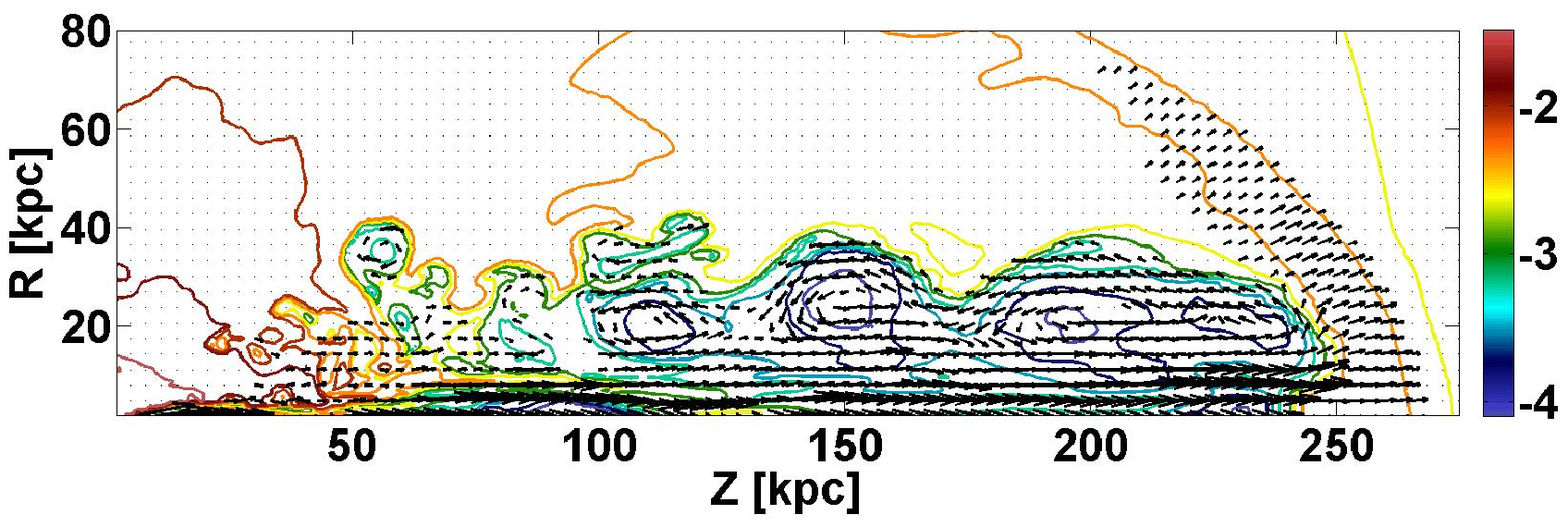}}
      \caption{Results of the standard run at $t=100 \Myr$. Axes are in kpc. The horizontal boundary {{{(labelled $Z$)}}} is the symmetry axis of the two jets. {{{ The left vertical boundary of the figure (labelled $R$) is taken along the mirror-symmetry plane of the flow $z=0$; it is termed the equatorial plane.}}}	
      The total two jets power is $P_{2j}= 4 \times 10^{45} \erg \s^{-1}$, the initial jet velocity is
      $v_j = 1.33 \times 10^{4} \km \s^{-1}$, the mass outflow from the two opposite jets (here
we calculate only one jet) is $\dot M_{2j} = 70 M_\odot \yr^{-1}$, and the jet half-opening angle is $\alpha=30^{\circ}$.
      The jet is injected at $r=1 \kpc$, a region that is not well resolved in the figure. We mark several features that are discussed in the text.
       \subref{subfigure:stdt} The temperature map of the standard run in logarithmic scale and units of K.
       \subref{subfigure:stdne} The electron density in logarithmic scale and units of $\cm^{-3}$.
       \subref{subfigure:stdvortices} Contours of the electron density in logarithmic scale and velocity arrows.
       The velocity arrows are divided into 4 velocity bins: $500-1000 \km \s^{-1}$, $1000-5000 \km \s^{-1}$, $5000-10^4 \km \s^{-1}$
       and above $10^4 \km \s^{-1}$ -  from shortest to longest. Velocities below $500 \km \s^{-1}$ do not appear in the figure.
       The fragmented vortices, that are at the heart of the present study, are marked.  Vortex shedding region is also marked.}
      \label{fig:standard}
\end{figure}

In figure \ref{fig:standard} we present our standard run at time $t=100 \Myr$.
We focus our study on the vortices that can be clearly seen along the shocked jet's material, i.e., the cocoon.
Between the large vortices we can see denser ICM gas segments flowing toward the axis.
In our simulations the axi-symmetry implies that each such segment is actually a torus.
These segments are marked by ``KH instability''.
There is a bow shock running into the ICM, the forward shock.
Most of the kinetic energy of the jet is dissipated in the reverse shock, as marked on the figure.
There is a contact discontinuity surface between the shocked ICM and the shocked jet material.
It can be located where the large temperature and density gradients are.

\begin{figure}
  \centering
 \subfigure{\includegraphics[scale=0.4,clip=true,trim=40 0 50 0]{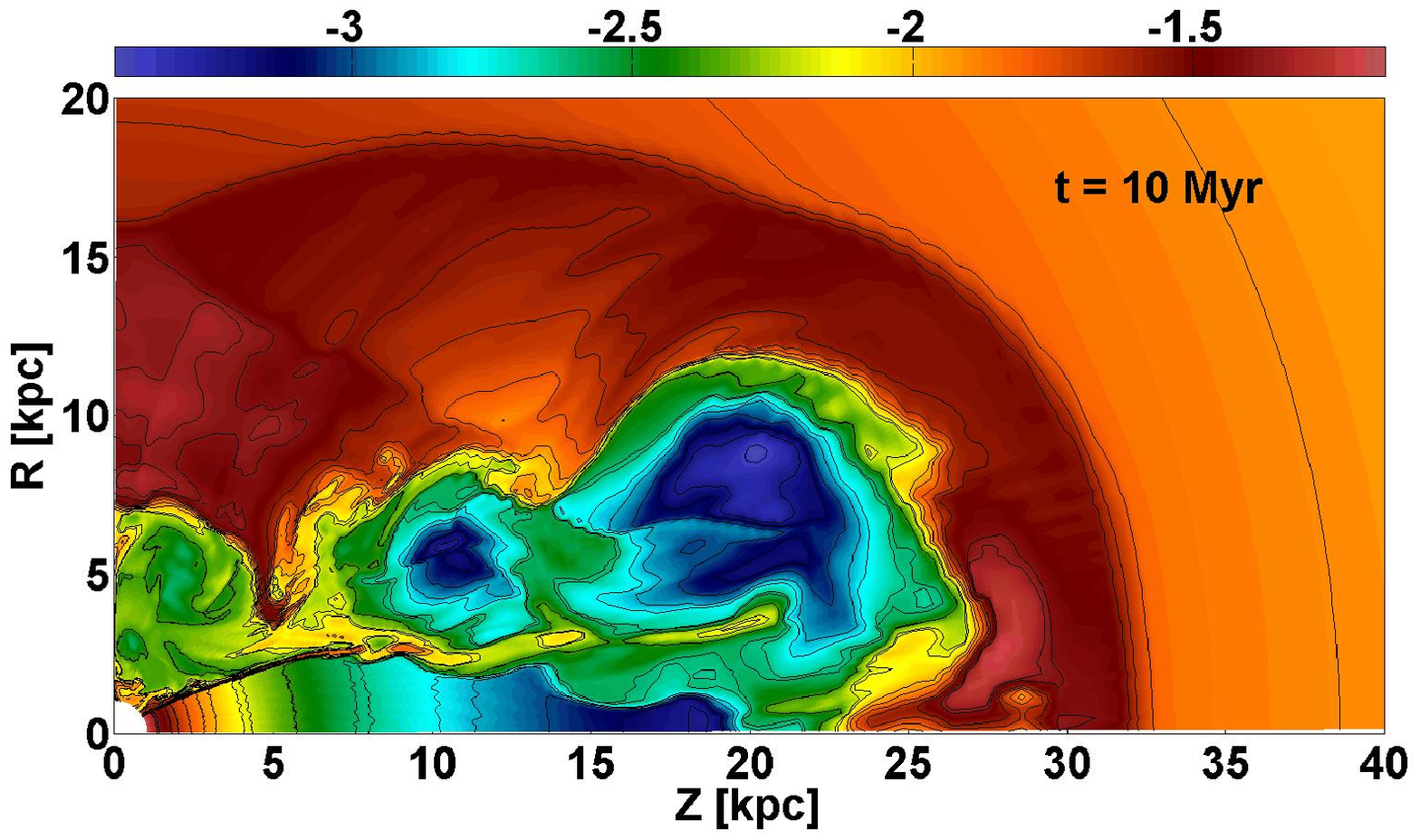}}
 \subfigure{\includegraphics[scale=0.4,clip=true,trim=40 0 50 0]{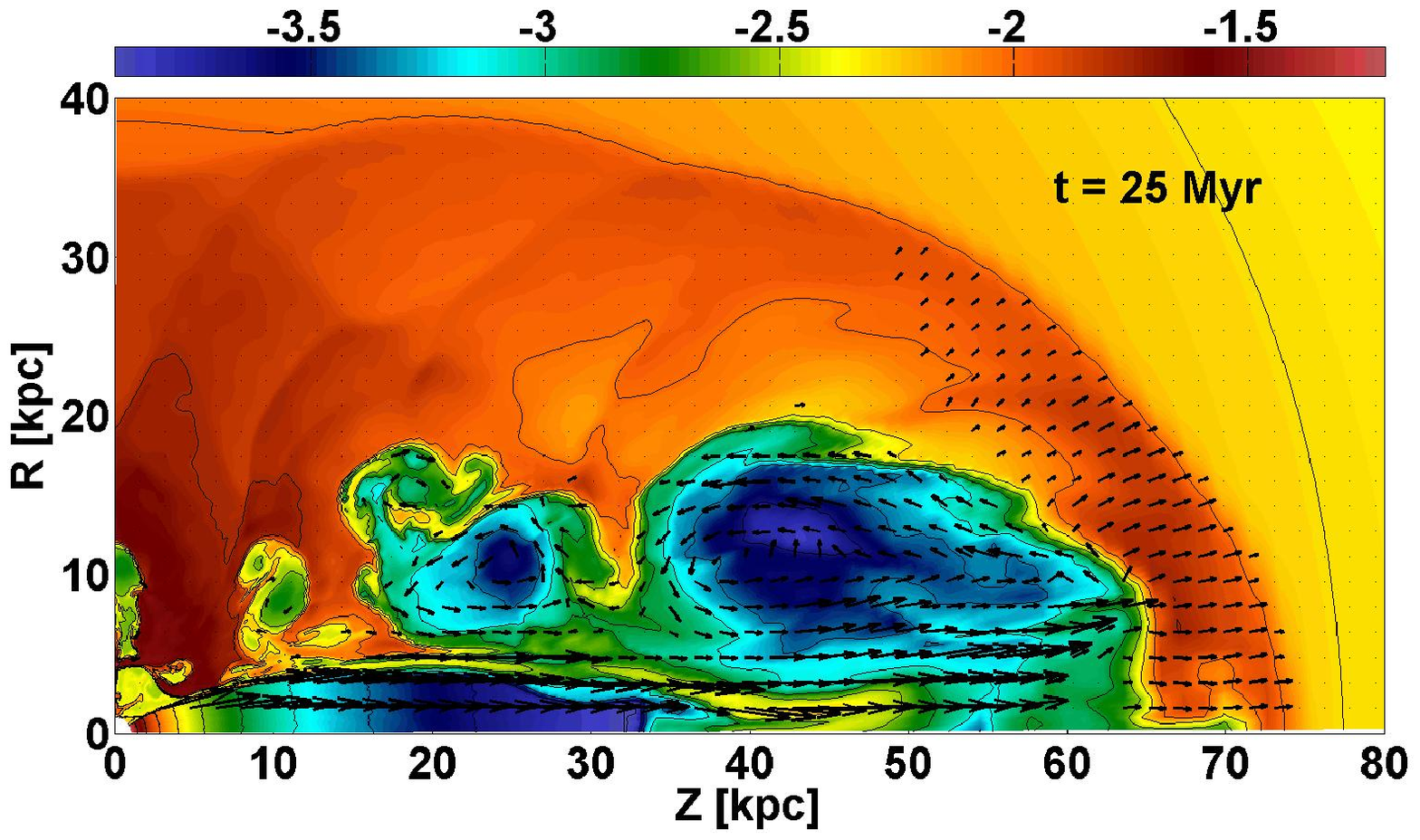}}\\
 \subfigure{\includegraphics[scale=0.4,clip=true,trim=40 0 50 0]{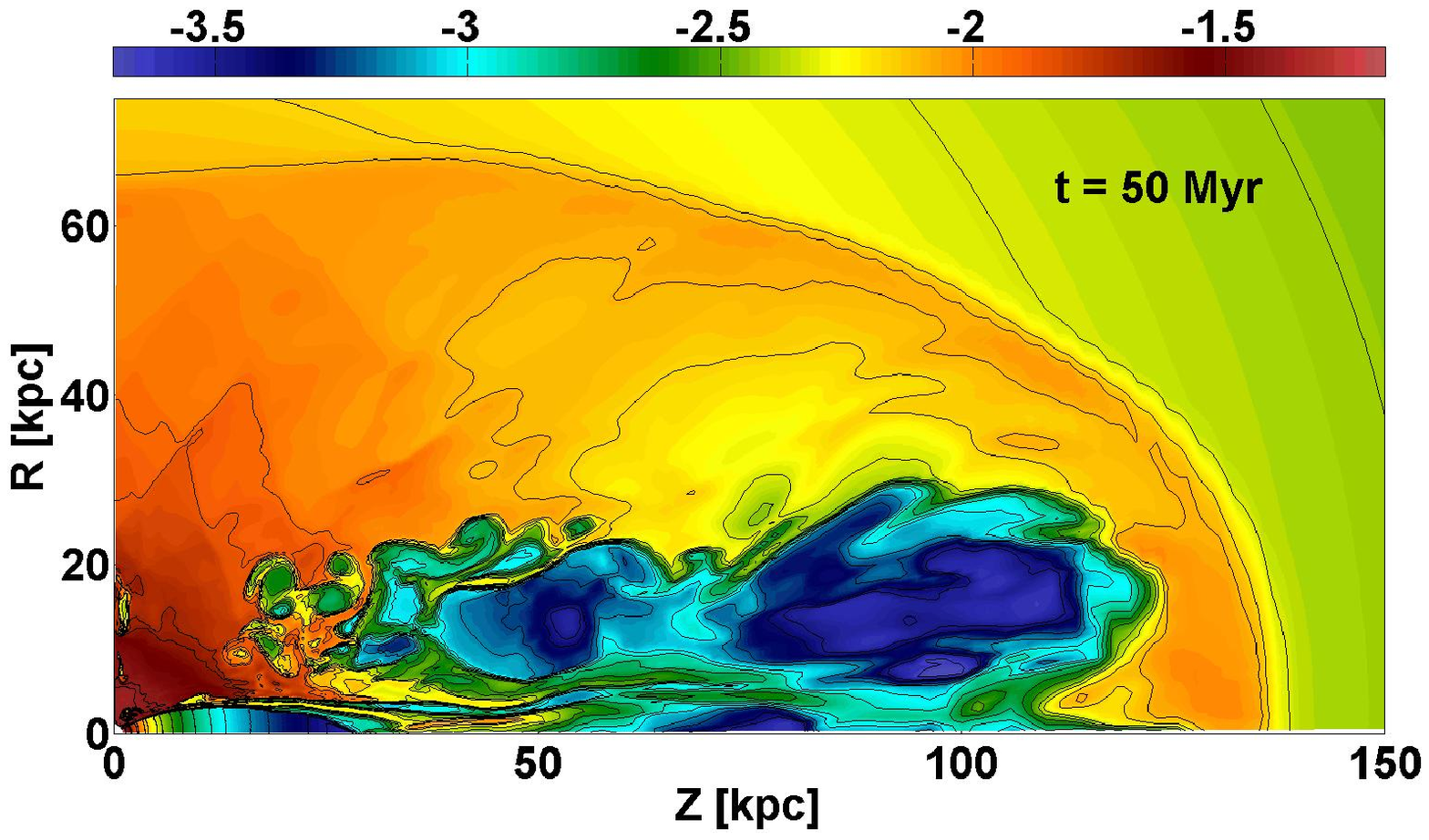}}
 \subfigure{\includegraphics[scale=0.4,clip=true,trim=40 0 50 0]{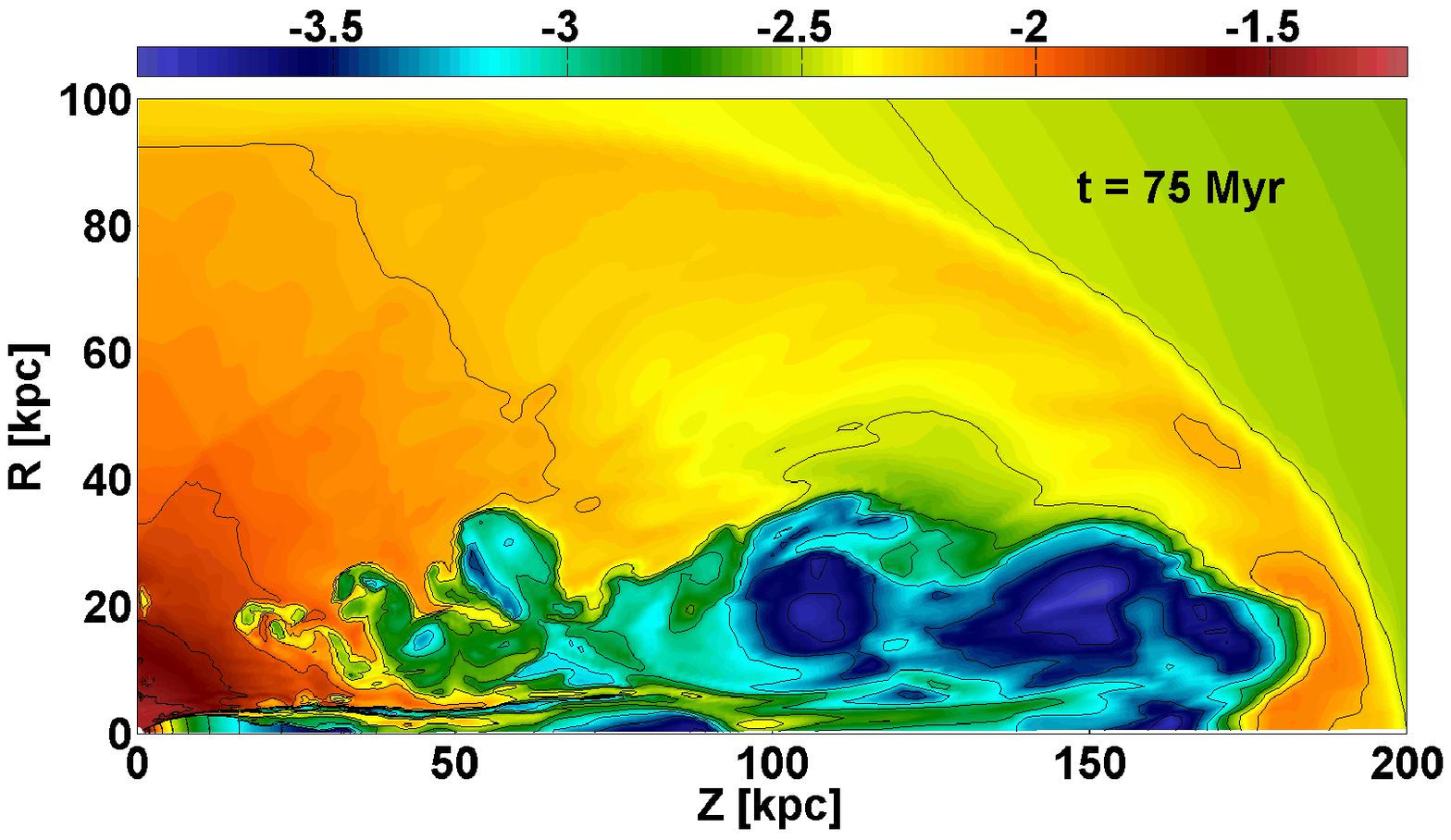}}
\caption{Density maps of the standard run at four times. The color scaling is of $\log n_e({\rm cm}^{-3}$),
 where $n_e$ is the electron density. A primary vortex right behind the jet's head and another vortex trailing it
  exist during the evolution. At the $t=10\Myr$ picture the flapping of the jet may be seen as a perturbation on the
  separating layer between the expanding jet and the cocoon. }
  \label{fig:stdevolution}
\end{figure}

The primary source of vorticity in the simulation is the back-flow of the
shocked jet material reflected from the shock front.
As may be seen in figure \ref{fig:stdevolution}, a large vortex immediately behind the jet's head, the ``primary vortex'', exists at all times,
and it creates a cascade of smaller vortices.

Morphologically, these vortices can be identified with bubbles (see also \citealt{Sternberg2008b}).
Several process, as we now detail, influence the evolution of these vortices, and hence the bubbles'
morphology.
\begin{itemize}
\item {\it Kelvin-Helmholtz (KH) instability.} {{{ The  back-flow of the lower density cocoon material inside the vortex relative to the denser shocked ICM material forms a shear flow that is prone to the KH instability.}}}
The KH instability creates the wavy morphology, namely, it fragments large
bubbles to smaller ones.
\item {\it Rayleigh-Taylor (RT) instability.} The denser shocked ICM is on top of
the much lower density gas of the bubbles. This RT instability does not exist in the front of the
outer bubble \citep{Sternberg2008b}, but it might amplify the fragmentation formed by the KH instability in the tail.
\item {\it Vortex Shedding.} Vortex shedding occurs mainly on trailing vortices and
forms disconnected vortices \citep{Norman1996}.
\item {\it Jet flapping.} The jet is collimated by the pressure of the earlier shocked jet material and the shocked ICM.
Due to the instabilities listed above and the stochastic flow the collimating pressure is not constant.
As a result of that the jet's width changes. In 3D flow the jet will jitter around the symmetry axis.
This motion is termed flapping.
This flapping occurs in semi-periodic manner on a typical time scale of jet-width sound crossing time.
This flapping has two effects.
First, the jitter of the jet slows down the `drilling' of the jet through the ICM.
Second, the flapping seeds semi-periodic perturbation on the boundary of the primary bubble which
in turn amplified by the KH instability and at least at early stages, creates vortex-shedding.
The flapping is the subject of a future paper.
\end{itemize}

{{{ The type of flow simulated here is known to be KH-unstable, with formation of more than one vortex in the non-linear regime (e.g., \citealt{Norman1996, Mizuta2004, MG2012}). Our new emphasize is on the role of these large vortices in forming and determining the properties of observed X-ray deficient bubbles (cavities). Although magnetic fields might change the exact morphology of the vortices formed by the KH instability, they cannot suppress them completely. If the magnetic fields are entangled on small scales, as is expected in many cases (e.g., \citealt{Soker2010}), then the magnetic field can locally suppress small scale KH-instability modes. However, they cannot suppress the large scale modes that are considered here. 

In the opposite limit to small scale fields, the ICM magnetic field lines are stretched by the jet and exert tension on two opposite sides of the shocked jet's material (see, e.g.,  figure 3 in \citealt{Soker2010}).  If the magnetic fields are strong enough they can suppress even the large scale KH-instability modes, {\it but only along the field lines}. Namely, only in two opposite quarters around the jet. The two other quarters are prone to the KH instability and bubbles will be formed.  In any case, strong magnetic field that can suppress large scale KH-instability modes are not easy to obtain. \cite{MG2012}, for example, did not succeed in suppressing the KH instability in their MHD simulations.  }}}
  
\subsection{Chains of cavities}
\label{subsec:chain}

\begin{figure}
  \centering
  \subfigure[][]{\label{subfigure:xrayorig} \includegraphics[width=0.8\textwidth]{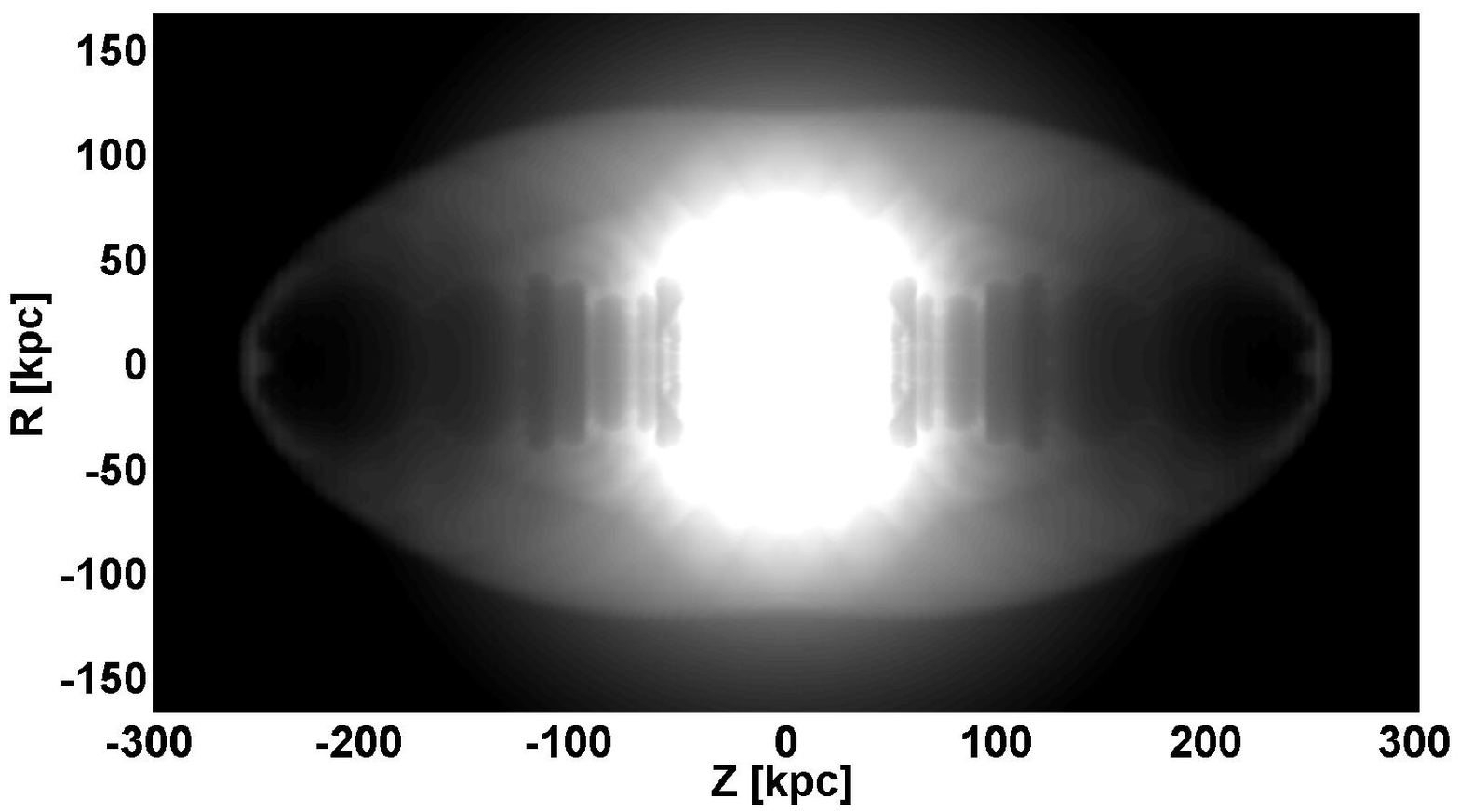}}\\
   \subfigure[][]{ \label{subfigure:xrayfilter}\includegraphics[width=0.8\textwidth]{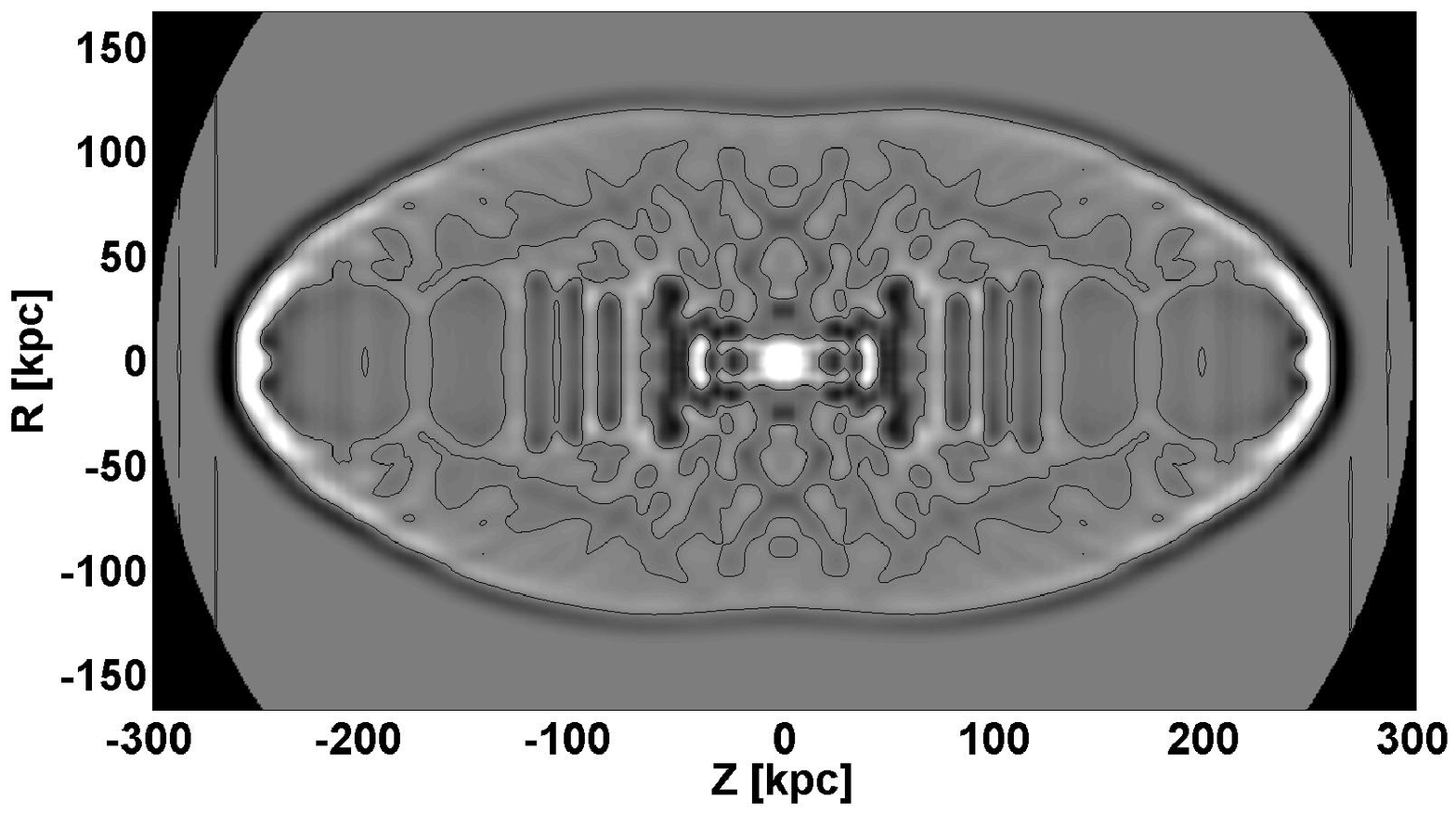}}
          \caption{Synthetic X-ray image of the standard simulation viewed from the equatorial plane direction (inclination of $90^\circ$) at $t=100 \Myr$, performed by integrating $n_e^2$ along the line of sight.
      \subref{subfigure:xrayorig} Original X-ray image.
      \subref{subfigure:xrayfilter} Unsharp filter applied to the image at \subref{subfigure:xrayorig}. Thin lines are the $1$ contours of the quotient image. The noisy contours which are not on the major axis are due to sound waves.
           }
      \label{fig:xray}
\end{figure}

The aforementioned mechanisms of vortex formation and
fragmentation lead to the formation of a chain of vortices along the jet axis (figure \ref{subfigure:stdvortices}).
To compare with observation we integrated the quantity $n_e^2$ along the line of sight perpendicular to the symmetry axis.
To a good approximation $n_e^2$ is proportional to the observed X-ray emission.
To emphasize the cavities we applied Gaussian unsharp filter to the integrated map, with filter parameters
of $\sigma_{\rm coarse}=30$ and $ \sigma_{\rm fine}=10$ \citep{Dong2010}.
The results are presented in figure \ref{fig:xray}.
We can identify two large bubbles and a few smaller ones closer to the center at each side of the equatorial plane. The center of the large bubbles are located approximately at coordinates $(r,z)=(0,\pm150)$ and $(0,\pm210)$. The denser medium between the large bubbles which creates a brighter X-ray emission and separates the bubbles is seen at $z=\pm130$ and $z=\pm175$.

Our results can account for the structure of two opposite chains of bubbles
that are close to each other, and even overlap, as observed for example
in Hydra A \citep{Wise2007} and in two bubbles of NGC 5813 \citep{Randall2011}.
The inner smaller bubbles that are seen in our simulations are likely to be smeared in observations due to departure from pure axi-symmetry, inclination, and low resolution, and appear as one bubble.
However, under different numerical parameters, more than two large bubbles on one side can be obtained.

We now emphasize two important points. First, we note that because of the highly turbulent nature of the flow we expect
the simulations not to be convergent in terms of fine structural details, i.e. small vortices and thin filaments.
Moreover, small differences in the initial parameters or numerical noise may result in macroscopic differences due to these instabilities.
Such differences may include the exact position of the forward shock, the number of macroscopic vortices and whether the jet
penetrates through the primary vortex.
The total volume of the bubbles is determined by energy considerations, and does not depend on the these instabilities.
As well, the general large-scale morphology does not depend on the instabilities.
Indeed, we find that chains of bubbles are formed for a non-negligible set of jet parameters (see section \ref{subsec:survey}).

Second, in order to appear like a chain of bubbles in the surface brightness map, the vortices shouldn't
necessarily be completely disconnected. It is sufficient that a relatively small ICM segment
starts to fragment a bubble into two regions, for the structure to appear like two adjacent bubbles in the X-ray image. In fact, in most our simulations the bubbles are at least partially connected by the jet.
Such dense ICM segments are regularly formed by vortices in our simulations, and hence a chain of bubbles is not a rare outcome.

\subsection{Parameter space survey}
\label{subsec:survey}
\begin{figure}
  \centering
 \subfigure[][$\alpha=15^{\circ}$, $v_j=6.5\times10^3\km \s^{-1} $]{\label{subfigure:survey_narrow1}\includegraphics*[scale=0.25,clip=true,trim=0 0 50 0]{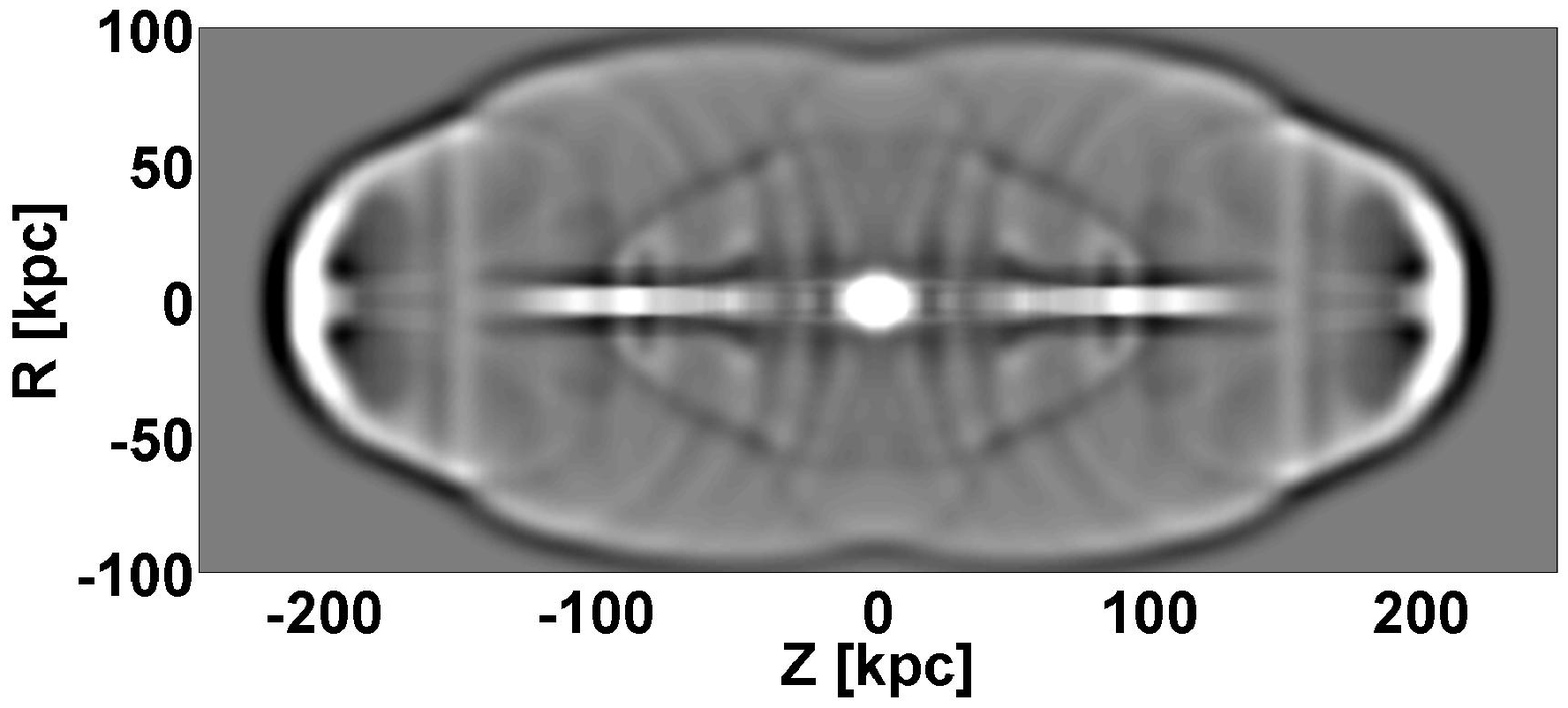}}
 \subfigure[][$\alpha=15^{\circ}$, $v_j=1.3\times10^4\km \s^{-1} $]{\label{subfigure:survey_narrow2}\includegraphics*[scale=0.25,clip=true,trim=0 0 50 0]{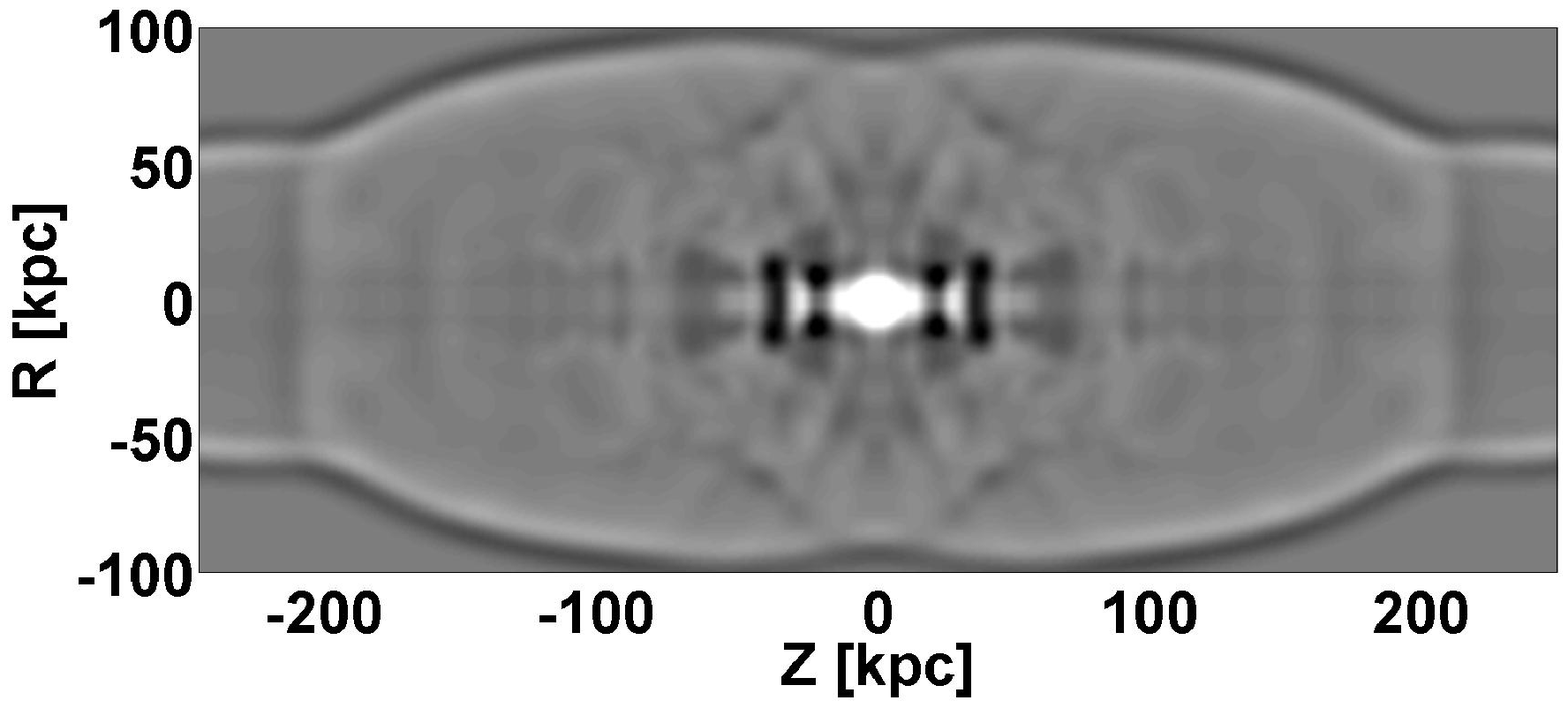}}
 \subfigure[][$\alpha=15^{\circ}$, $v_j=2.7\times10^4\km \s^{-1} $]{\label{subfigure:survey_narrow3}\includegraphics*[scale=0.25,clip=true,trim=0 0 50 0]{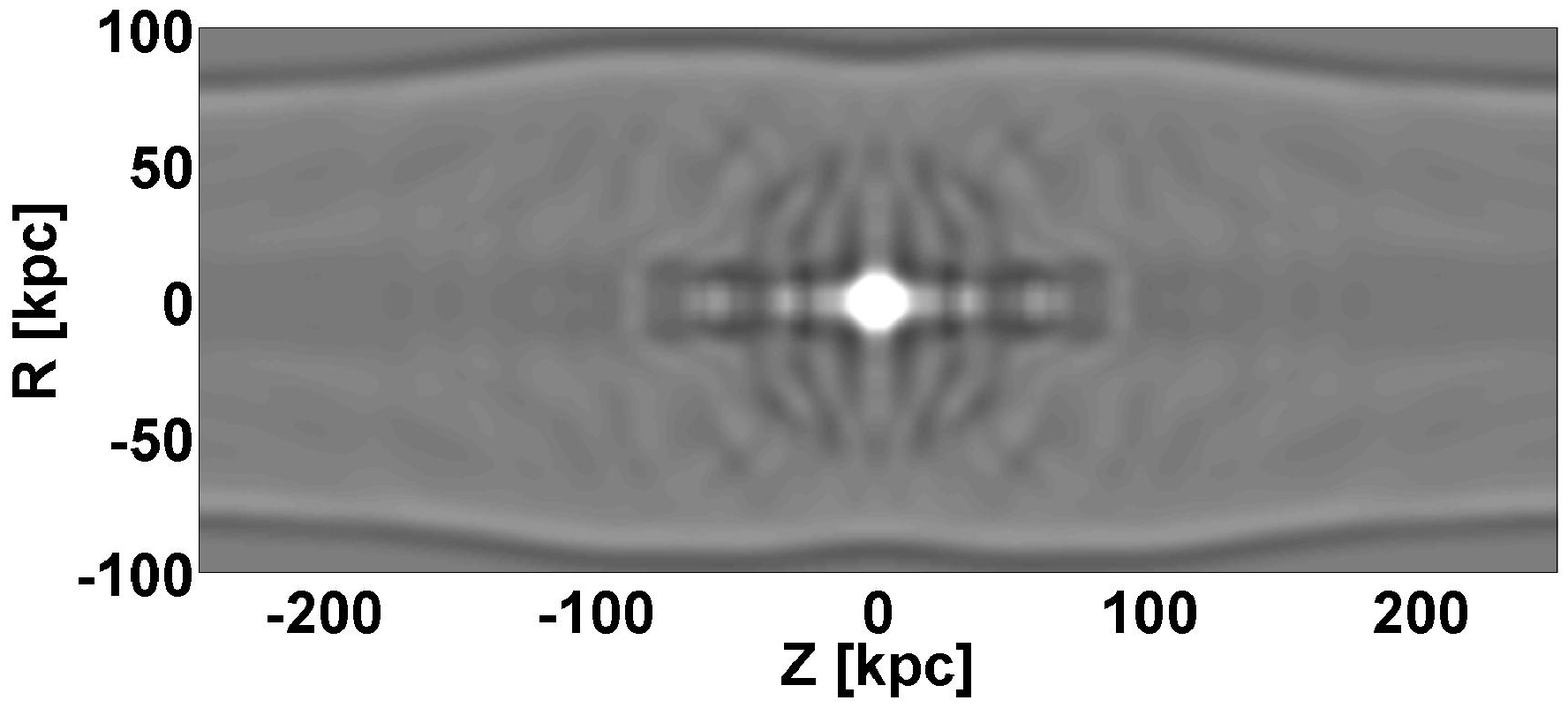}}\\
 \subfigure[][$\alpha=30^{\circ}$, $v_j=7\times10^3\km \s^{-1} $]{\label{subfigure:survey_mid1}\includegraphics*[scale=0.25,clip=true,trim=0 0 50 0]{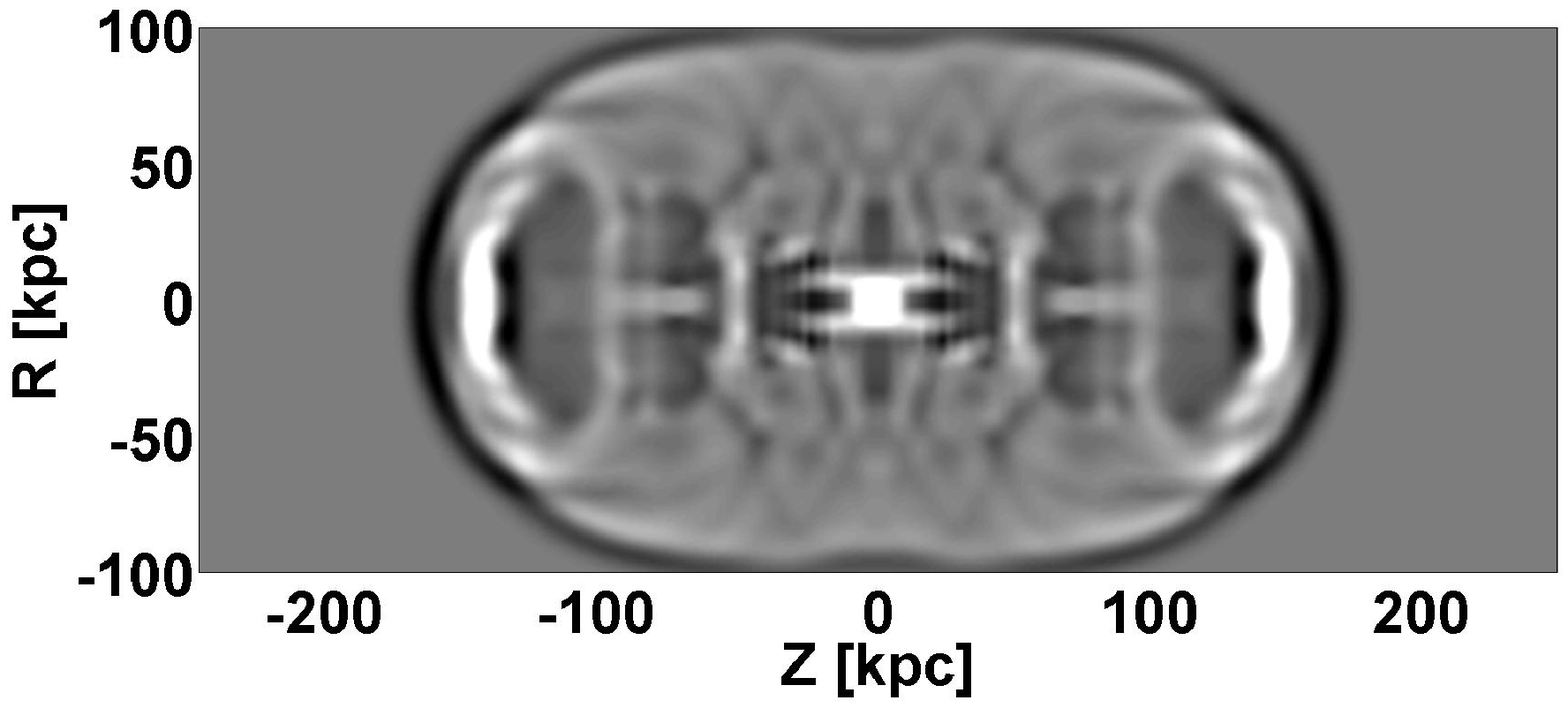}}
 \subfigure[][$\alpha=30^{\circ}$, $v_j=1.3\times10^4\km \s^{-1} $]{\label{subfigure:survey_stdrun}\includegraphics[scale=0.25,clip=true,trim=0 0 50 0]{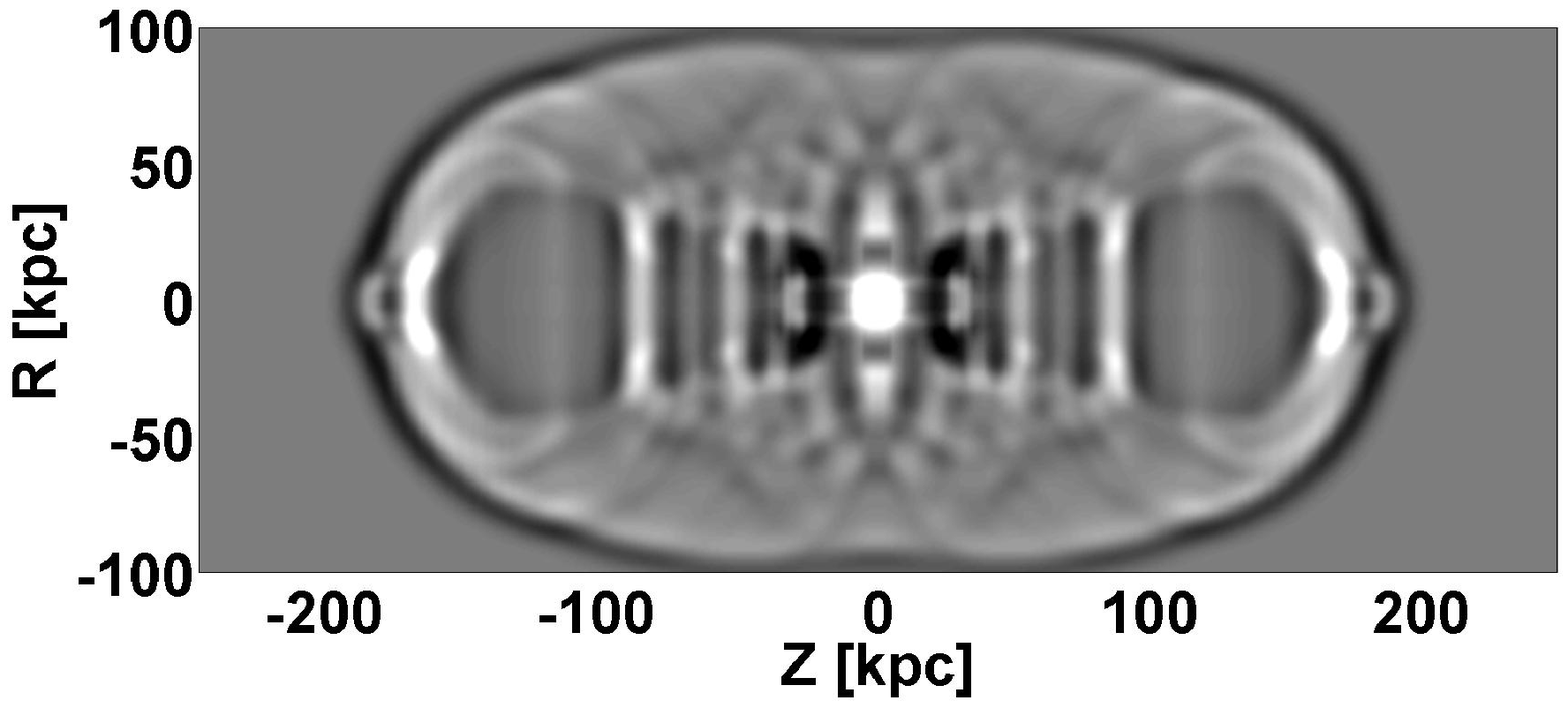}}
 \subfigure[][$\alpha=30^{\circ}$, $v_j=2.7\times10^4\km \s^{-1} $]{\label{subfigure:survey_mid2}\includegraphics*[scale=0.25,clip=true,trim=0 0 50 0]{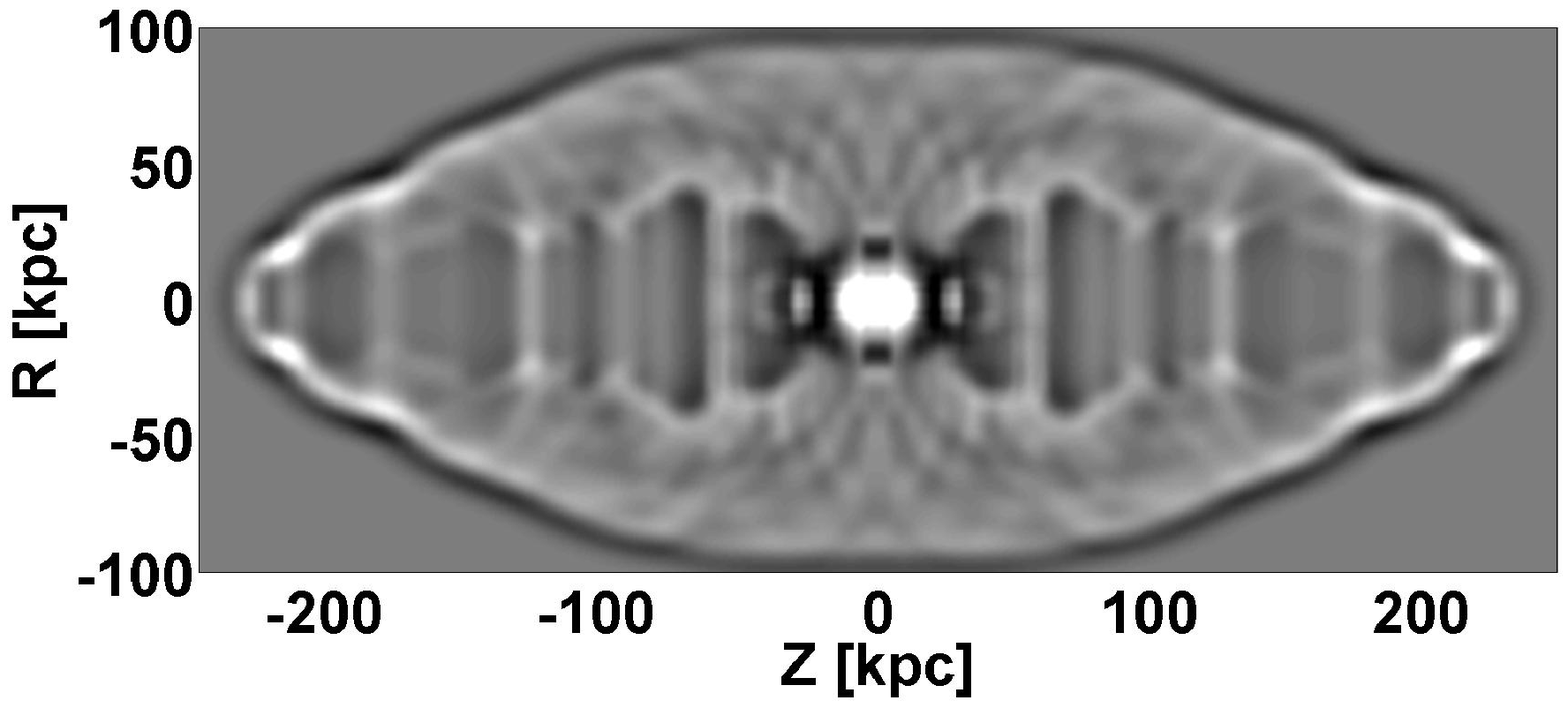}}\\
 \subfigure[][$\alpha=45^{\circ}$, $v_j=7\times10^3\km \s^{-1} $]{\label{subfigure:survey_wide1}\includegraphics*[scale=0.25,clip=true,trim=0 0 50 0]{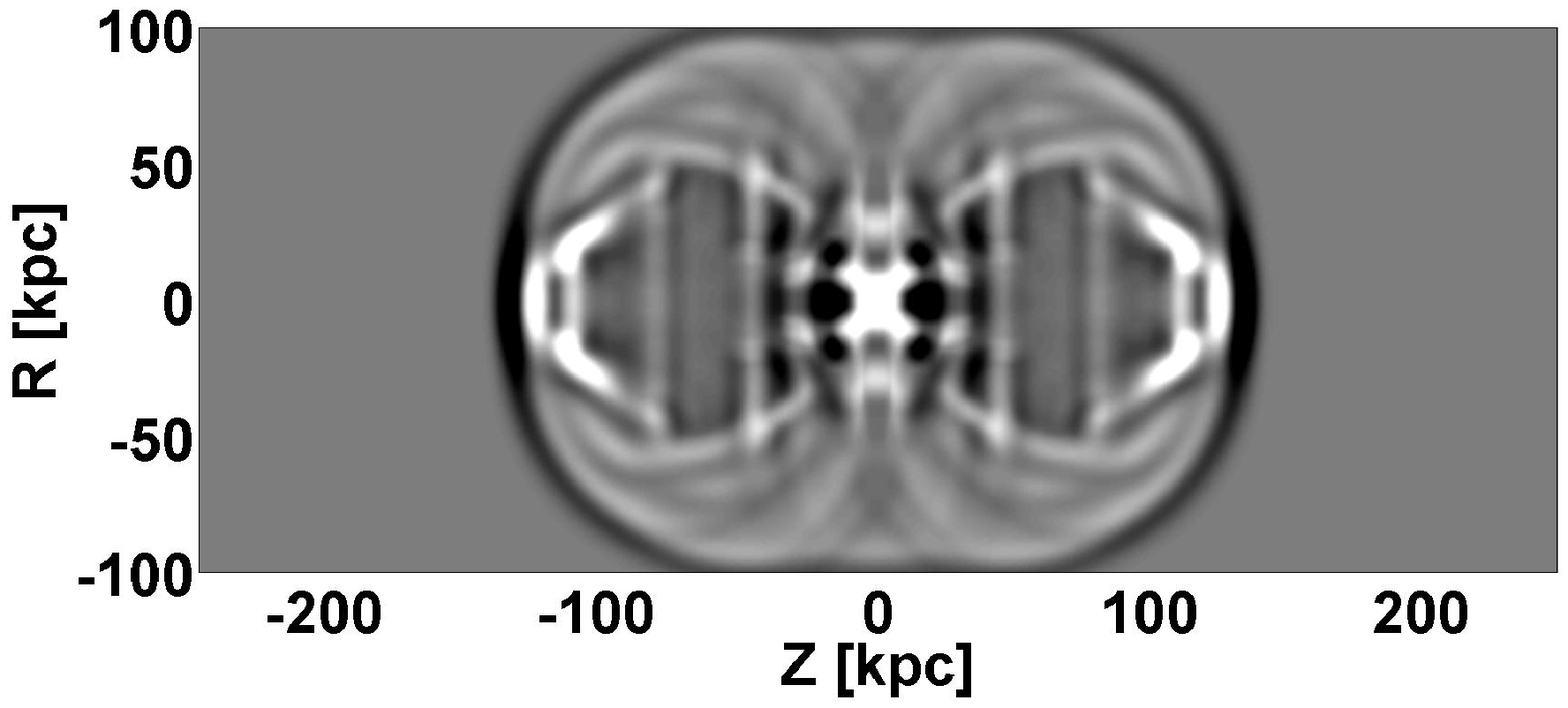}}
 \subfigure[][$\alpha=45^{\circ}$, $v_j=1.3\times10^4\km \s^{-1} $]{\label{subfigure:survey_wide2}\includegraphics*[scale=0.25,clip=true,trim=0 0 50 0]{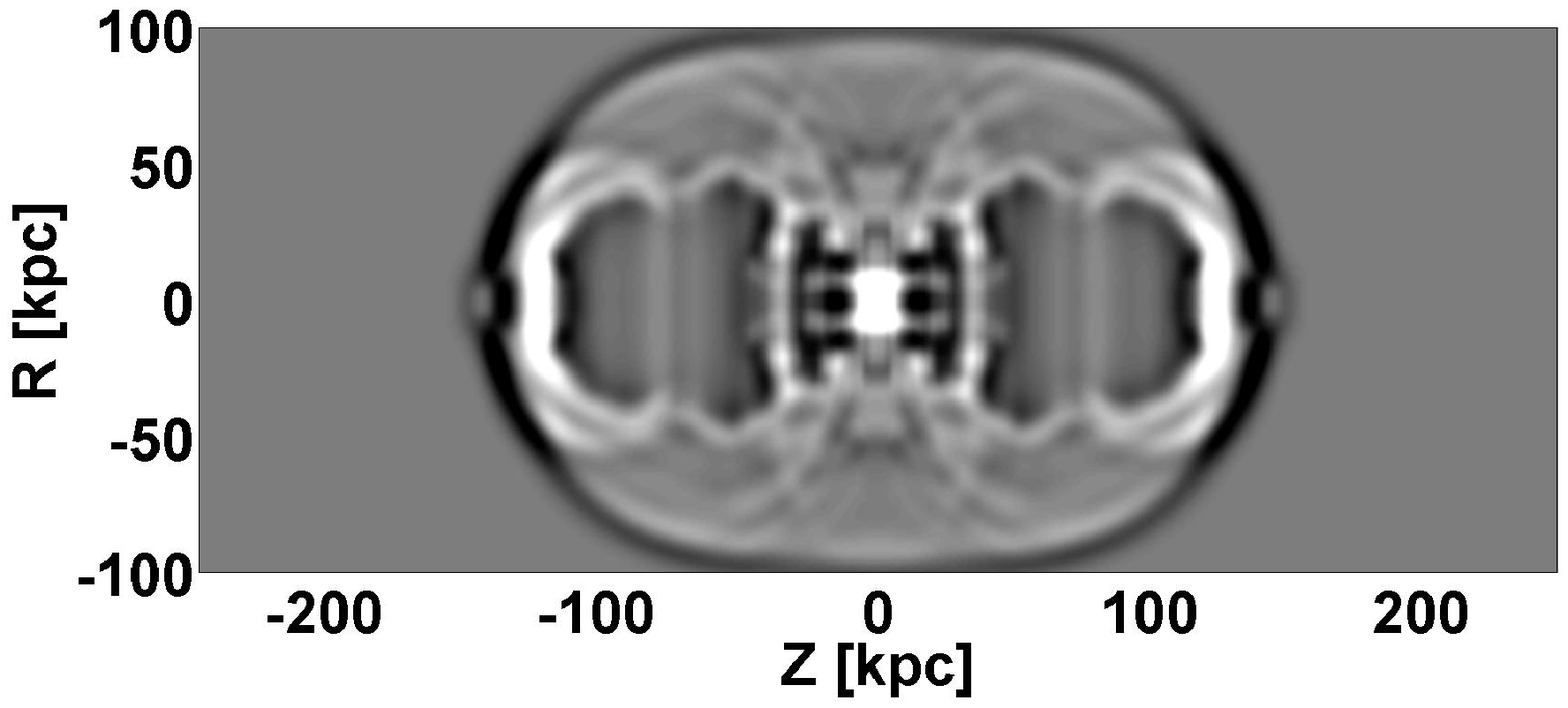}}
 \subfigure[][$\alpha=45^{\circ}$, $v_j=2.7\times10^4\km \s^{-1} $]{\label{subfigure:survey_wide_fast}\includegraphics*[scale=0.25,clip=true,trim=0 0 50 0]{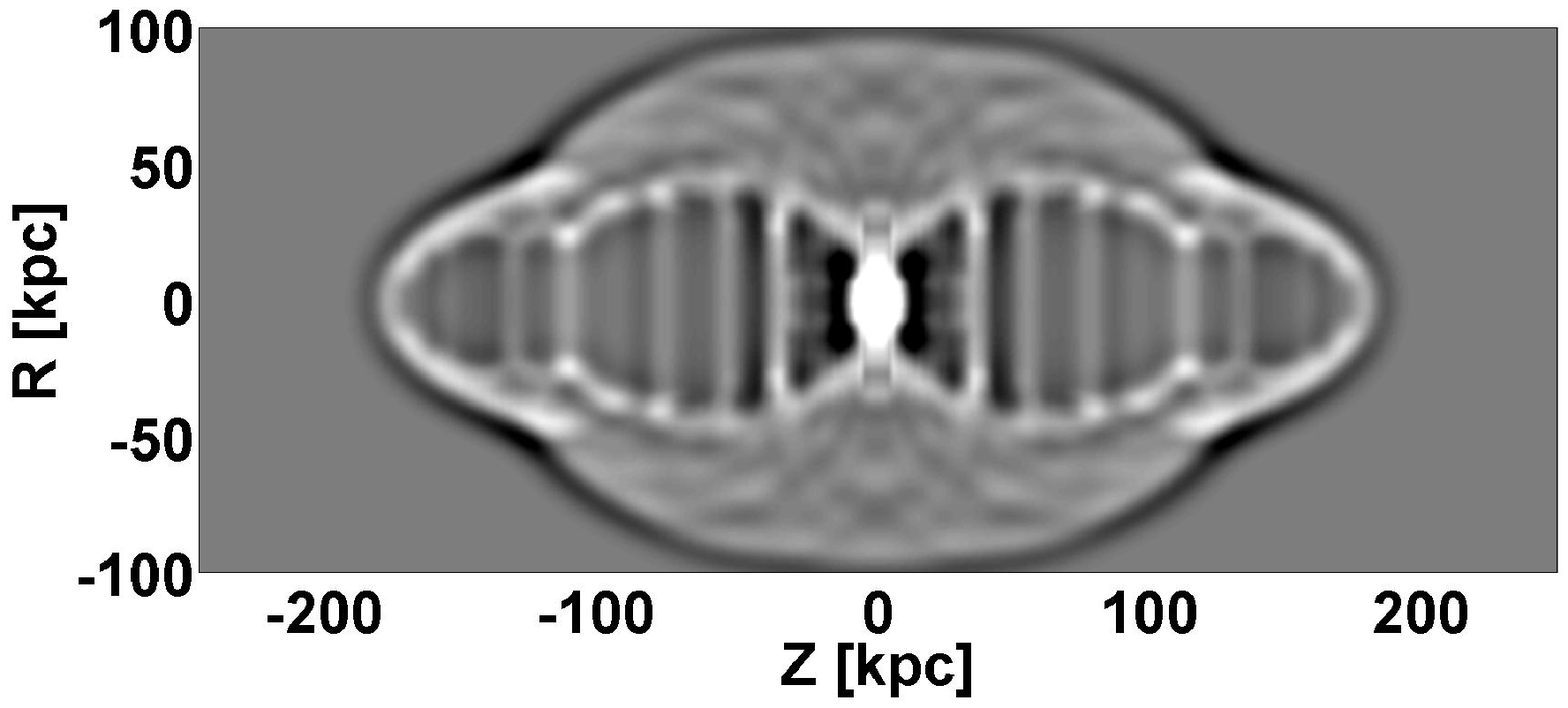}}
\caption{ Synthetic X-ray images as in figure \ref{fig:xray}, but for different values of the jet's half opening angle $\alpha$ and velocity $v_j$, as specified in each panel. Panel \subref{subfigure:survey_stdrun} is our standard case. The rest of the parameters are identical to those of the standard run (see section \ref{sec:numerical}). }
  \label{fig:survey}
\end{figure}

We examine the influence of the jet's half opening angle $\alpha$ and of its velocity $v_j$ on the appearance of the
chain of bubbles viewed perpendicular to the jets' axis. {{{ We have studied the range $15^{\circ} \le \alpha \le 60^{\circ}$ and 
$7\times10^3\km \s^{-1} \le v_j \le 3\times10^4\km \s^{-1}$.}}}
In figure \ref{fig:survey} we present the calculated X-ray map of a subset of the runs after applying the unsharp filter (as explained in section \ref{subsec:chain}).

In most of these synthetic images bright X-ray thin segments divide the X-ray deficient volume into several apparent bubbles.
These thin segments are the KH-filaments of the ICM pushing between the bubbles (the area marked by 'KH-instability' in figure \ref{subfigure:stdt}).
In reality, inclination and departure from pure axisymmetry will make the segments less prominent.
Therefore, in observations only cases with high segment to bubble contrast will appear like a chain of bubbles.
>From figure \ref{fig:survey} we find that these cases are typical for mid-range opening angles {{{($\alpha=30^{\circ} $, figures \ref{subfigure:survey_mid1},\ref{subfigure:survey_stdrun},\ref{subfigure:survey_mid2}), but with some dependence on the jet velocity.
In cases with slower jets ($v_j\le1.3\times10^4\km \s^{-1}$) and larger opening  angles ($ \alpha\ge 45^{\circ}$) as in figures \ref{subfigure:survey_wide1} and \ref{subfigure:survey_wide2}, they will mostly appear like a single bubble.}}}
For very fast and narrow jets, the jet escapes without creating bubbles {{{(figures \ref{subfigure:survey_narrow1},\ref{subfigure:survey_narrow2},\ref{subfigure:survey_narrow3})}}}.

\section{SUMMARY}
\label{sec:summary}

We showed that a combination of various vortex fragmentation mechanisms, mainly KH-instability and vortex-shedding,
can explain chains of close and overlapping X-ray cavities (bubbles) observed in clusters of galaxies.
We used the hydrodynamic PLUTO code \citep{Mignone2007} in 2.5D, i.e. spherical coordinate system with cylindrical symmetry.
The general flow structure is presented in figure \ref{fig:standard}, where two large bubbles can clearly be identified.
A primary (front) vortex exists during the entire simulation with one or two large trailing vortices and smaller vortices shedded further downstream (figure \ref{fig:stdevolution}).
The large vortices are interpreted as the X-ray deficient bubbles seen in observations, as demonstrated in figure \ref{fig:xray}.
Although chain of bubbles are formed in a wide range of parameters, for some parameters, like for wide jets,   the bright X-ray segments separating bubbles are too faint and only one bubble is formed for each jet launching episode. This is demonstrated in figure \ref{fig:survey}.

We find the formation of many vortices by the KH-instability and the vortex shedding process to be
very interesting and significant in determining the structure and evolution of bubbles.
It may also play an important role by dredging material up from the
inner regions outward and by dissipating heat to the inner region \citep{Gilkis2012}.
In a future study we will explore the entrainment and lifting of ICM by the chain of vortices, a process
that is observed to occur in Hydra A \citep{Kirkpatrick2009}.

Basically we have shown that one jet activity episode can form two opposite chains of close bubbles
as observed for example in Hydra A \citep{Wise2007} and possibly in NGC 5813 \citep{Randall2011}.
Large separations between bubbles, on the other hand, probably require multiple jet activity episodes.

{{{We thank an anonymous referee for helpful comments.}}} This research was supported by the Asher Fund for Space Research at the Technion, and the Israel Science foundation.


\begin{thebibliography}{999999999999999999999999999999}                                                                   %


\bibitem[Alouani Bibi et al.(2007)]{AlouaniBibi2007}Alouani Bibi, F., Binney,
J., Blundell, K., \& Omma, H.\ 2007, \apss, 311, 317

\bibitem[Baldi et al.(2009)]{Baldi2009}Baldi, A., Forman, W., Jones, C.,
Kraft, R., Nulsen, P. Churazov, E., David, L., Giacintucci, S. \ 2009, \apj,
707, 1034

\bibitem[Basson \& Alexander(2003)]{Basson2003}Basson, J.~F., \& Alexander,
P.\ 2003, \mnras, 339, 353

\bibitem[Binney \& Tabor(1995)]{Binney1995}Binney, J., \& Tabor, G.\ 1995,
\mnras, 276, 663

\bibitem[Birzan et al.(2011)]{Birzan2011}Birzan, L., Rafferty, D.~A.,
McNamara, B.~R., Nulsen, P.~E.~J., \& Wise, M.~W.\ 2011, \memsai, 82, 573

\bibitem[Blanton et al.(2011)]{Blanton2011}Blanton, E.~L., Randall, S.~W.,
Clarke, T.~E., Sarazin, C. L., McNamara, B. R., Douglass, E. M., McDonald, M.
2011, \apj, 737, 99

\bibitem[Br{\"u}ggen et al.(2007)]{Bruggen2007}Br{\"u}ggen, M., Heinz, S.,
Roediger, E., Ruszkowski, M., \& Simionescu, A.\ 2007, \mnras, 380, L67

\bibitem[Cavagnolo et al.(2011)]{Cavagnolo2011}Cavagnolo, K.~W., McNamara,
B.~R., Wise, M.~W., et al.\ 2011, \apj, 732, 71

\bibitem[David et al.(2011)]{David2011}David, L.~P., et al. \ 2011, \apj, 728, 162

\bibitem[David et al.(2009)]{David2009}David, L.~P., Jones, C., Forman, W.,
     Nulsen, P.~E.~J., Vrtilek, J., O'Sullivan, E., Giacintucci, S., Raychaudhury, S. 2009, \apj, 705, 624

\bibitem[Dong et al.(2010)]{Dong2010}Dong, R., Rasmussen, J., \& Mulchaey,
J.~S.\ 2010, \apj, 712, 883

\bibitem[Doria et al.(2012)]{Doria2012} Doria, A., Gitti, M.,  Ettori, S.,
et al.\ 2012, arXiv:1204.6191

\bibitem[Fabian et al.(2011)]{Fabian2011}Fabian, A.~C., et al.\ 2011, \mnras,
418, 2154

\bibitem[Falceta-Goncalves et al.(2010)]{Falceta-Goncalves2010}%
Falceta-Goncalves, D., Caproni, A., Abraham, Z., Teixeira, D. M., \& de
Gouveia Dal Pino, E. M. 2010, ApJ, 713, L74

\bibitem[Farage et al.(2012)]{Farage2012}Farage, C.~L., McGregor, P.~J., \&
Dopita, M.~A. 2012, \apj, 747, 28

\bibitem[Friedman et al.(2012)]{Friedman2012}Friedman, S.~H., Heinz, S., \&
Churazov, E.\ 2012, \apj, 746, 112

\bibitem[Gaspari et al.(2012a)]{Gaspari2012a}Gaspari, M., Brighenti, F., \&
Temi, P.\ 2012, arXiv:1202.6054

\bibitem[Gaspari et al.(2012b)]{Gaspari2012b}Gaspari, M., Ruszkowski, M., \&
Sharma, P.\ 2012, \apj, 746, 94

\bibitem[Gastaldello et al.(2009)]{Gastaldello2009}Gastaldello, F., Buote,
D.~A., Temi, P., et al.\ 2009, \apj, 693, 43

\bibitem[Gilkis \& Soker(2012)]{Gilkis2012}Gilkis, A., \& Soker, N.\ 2012, posted on astro-ph

\bibitem[Gitti et al.(2012)]{Gitti2012}Gitti, M., Brighenti, F., \& McNamara,
B.~R.\ 2012, Advances in Astronomy, 2012,

\bibitem[Gitti et al.(2010)]{Gitti2010}Gitti, M., O'Sullivan, E., Giacintucci,
S., David, L. P.; Vrtilek, J., Raychaudhury, S., Nulsen, P. E. J. \ 2010,
\apj, 714, 758

\bibitem[Heinz et al.(2006)]{Heinz2006}Heinz, S., Br{\"u}ggen, M., Young, A.,
\& Levesque, E.\ 2006, \mnras, 373, L65

\bibitem[Kirkpatrick et al.(2009)]{Kirkpatrick2009}Kirkpatrick, C.~C., Gitti,
M., Cavagnolo, K.~W., McNamara, B. R., David, L. P., Nulsen, P. E. J., Wise,
M. W.\ 2009, \apjl, 707, L69
{{{
\bibitem[Mathews \& Guo(2012)]{MG2012}Mathews, W.~G., \& Guo, F.\ 2012, arXiv:1206.4997
}}}

\bibitem[McNamara et al.(2000)]{McNamara2000}McNamara, B.~R., Wise, M., Nulsen, P.~E.~J., et al.\ 2000,
\apjl, 534, L135

\bibitem[Mendygral et al.(2012)]{Mendygral2012}Mendygral, P., Jones, T., \&
Dolag, K.\ 2012, arXiv:1203.2312

\bibitem[Mendygral et al.(2011)]{Mendygral2011}Mendygral, P.~J., O'Neill,
S.~M., \& Jones, T.~W.\ 2011, \apj, 730, 100

\bibitem[Mignone et al.(2007)]{Mignone2007}Mignone, A., Bodo, G., Massaglia,
S., Matsakos, T., Tesileanu, O., Zanni, C., Ferrari, A., 2007, \apjs, 170, 228

\bibitem[Mizuta et al.(2004)]{Mizuta2004}Mizuta, A., Yamada, S., 
\& Takabe, H.\ 2004, \apj, 606, 804

\bibitem[Morsony et al.(2010)]{Morsony2010}Morsony, B.~J., Heinz, S.,
Br{\"u}ggen, M., \& Ruszkowski, M.\ 2010, \mnras, 407, 1277

\bibitem[Norman(1996)]{Norman1996}Norman, M.~L.\ 1996, Energy
Transport in Radio Galaxies and Quasars, 100, 319

\bibitem[Omma et al.(2004)]{Omma2004}Omma, H., Binney, J., Bryan, G., \& Slyz,
A.\ 2004, \mnras, 348, 1105

\bibitem[O'Neill \& Jones(2010)]{ONeill2010}O'Neill, S.~M., \& Jones,
T.~W.\ 2010, \apj, 710, 180

\bibitem[O'Sullivan et al.(2011)]{OSullivan2011}O'Sullivan, E., Giacintucci,
S., David, L.~P., Gitti, M., Vrtilek, J. M., Raychaudhury, S., Ponman, T.
J.\ 2011, \apj, 735, 11

\bibitem[Pandge et al.(2012)]{Pandge2012}Pandge, M.~B., Vagshette, N.~D.,
David, L.~P., \& Patil, M.~K.\ 2012, arXiv:1202.1364

\bibitem[Randall et al.(2011)]{Randall2011}Randall, S.~W., et al. \ 2011,
\apj, 726, 86

\bibitem[Roediger et al.(2007)]{Roediger2007}Roediger, E., Br{\"u}ggen, M.,
Rebusco, P., B{\"o}hringer, H., \& Churazov, E.\ 2007, \mnras, 375, 15

\bibitem[Soker(2009)]{Soker2009}Soker, N.\ 2009, \mnras, 398, L41
{{{
\bibitem[Soker(2010)]{Soker2010}Soker, N.\ 2010, arXiv:1007.2249
}}}
\bibitem[Sternberg et al. (2007)]{Sternberg2007}Sternberg, A., Pizzolato, F.
\& Soker N. 2007, \apj, 656, L5

\bibitem[Sternberg \& Soker(2008a)]{Sternberg2008a}Sternberg, A., \& Soker N.
2008a, \mnras, 384, 1327


\bibitem[Sternberg \& Soker(2008b)]{Sternberg2008b}Sternberg, A., \& Soker,
N.\ 2008b, \mnras, 389, L13


\bibitem[Sternberg \& Soker(2009)]{Sternberg2009}Sternberg, A., \& Soker,
N.\ 2009, \mnras, 395, 228


\bibitem[Vernaleo \& Reynolds(2006)]{Vernaleo2006}Vernaleo, J.~C., \&
Reynolds, C.~S.\ 2006, \apj, 645, 83

\bibitem[Wise et al.(2007)]{Wise2007}Wise, M.~W., McNamara, B.~R., Nulsen,
P.~E.~J., Houck, J.~C., \& David, L.~P.\ 2007, \apj, 659, 1153



\end{thebibliography}
\end{document}